\documentclass[12pt,a4paper]{article}

\usepackage{latexsym,amssymb,amsmath,amsbsy,epsfig}

\newcommand{\be}{\begin{equation}}
\newcommand{\ee}{\end{equation}}
\newcommand{\ba}{\begin{eqnarray}}
\newcommand{\ea}{\end{eqnarray}}
\newcommand{\bs}{\begin{subequations}}
\newcommand{\es}{\end{subequations}}
\newcommand{\no}{\nonumber\\}

\begin{document}

\title{
\normalsize \hfill CFTP/15-01
\\*[7mm]
\LARGE New texture-zero models for lepton mixing}

\author{
L.\ Lavoura\thanks{E-mail: {\tt balio@cftp.tecnico.ulisboa.pt}} \\
\small CFTP, Instituto Superior T\'ecnico, Universidade de Lisboa \\
\small Avenida Rovisco Pais 1, 1049-001 Lisboa, Portugal
}

\date{10 February 2015}

\maketitle

\begin{abstract}
I systematically consider,
in the context of the type-I see-saw mechanism,
all the predictive cases in which both the Dirac mass matrix
connecting the left-handed neutrinos to the right-handed neutrinos,
and the Majorana mass matrix of the latter neutrinos,
feature texture zeros,
while the mass matrix of the charged leptons is diagonal.
I have found a few cases
which had not been discussed in the literature previously.
\end{abstract}

\section{Introduction}

There has recently been renewed interest in the possibility of `texture' zeros
in the fermion mass matrices~\cite{Ludl:2014axa,Ferreira:2014vna,Ludl:2015lta}.
Texture zeros are well grounded in renormalizable field theories,
since they can always be implemented
through models with suitable Abelian symmetries and
(possibly many)
scalar fields with vacuum expectation values~\cite{Grimus:2004hf}.
The revival of interest was partially motivated
by the realization of the fact that a popular alternative approach,
where lepton mixing
is \emph{completely determined}\/ by a non-Abelian symmetry,
seems to have been fully explored~\cite{Fonseca:2014koa}.

With three Majorana neutrinos,
the mass Lagrangian is
\be
\mathcal{L}_\mathrm{mass} =
- \bar \ell_L M_\ell \ell_R
- \bar \ell_R M_\ell^\dagger \ell_L
+ \frac{1}{2} \left( \nu^T C^{-1} M \nu
- \bar \nu M^\ast C \bar \nu^T \right),
\ee
where $C$ is the charge-conjugation matrix in Dirac space.
The column-vector $\nu$ contains the three left-handed light-neutrino fields.
The neutrino Majorana mass matrix $M$ acts in flavour space and is symmetric.

In ref.~\cite{Frampton:2002yf},
$M_\ell$ was assumed to be diagonal
while $M$ had two zero matrix elements.\footnote{The matrix $M$
is $3 \times 3$ symmetric and therefore it has,
in general,
six independent matrix elements.
We say that ``$M$ has $n$ zero matrix elements''
if $n$ out of those six \emph{independent}\/ matrix elements vanish.
The actual total number of zero entries in $M$ will be larger than $n$
if some of the vanishing entries are off-diagonal.}
This was later generalized to the situation wherein $M_\ell$ is diagonal
and $M^{-1}$ has two zero matrix elements~\cite{Lavoura:2004tu};
mixed situations in which $M$ and $M^{-1}$ have one zero matrix element each,
while $M_\ell$ remains diagonal,
were studied in ref.~\cite{Dev:2010if}.
Recently,
all the cases in which both $M_\ell$ and $M$ sport texture zeros
were mapped~\cite{Ludl:2014axa,Ferreira:2014vna}.

It is known that the three light neutrinos are \emph{exceedingly}\/ light;
one of them may actually be massless.
Possibly the most popular theory for explaining that extreme lightness
is the (type~I) see-saw mechanism.
In that theory,
$M$ is not really a fundamental mass matrix,
rather
\be
M = - M_D M_R^{-1} M_D^T
\ee
is just the effective (approximate) mass matrix for the light neutrinos
arising out of the Dirac mass matrix $M_D$
connecting the standard neutrinos to some gauge-singlet
(``right-handed'') neutrino fields
and of the Majorana mass matrix $M_R$ of the latter.\footnote{We shall assume
in this paper that the number of right-handed neutrinos is three.}
In this context,
assuming the presence of texture zeros in $M$ seems unwarranted;
one should rather consider texture zeros in $M_D$ and $M_R$.
Indeed,
that was the rationale for ref.~\cite{Lavoura:2004tu},
where $M_D$ was assumed to be diagonal
(which means that it has six texture zeros)
and two texture zeros were enforced in $M_R$.

In this paper I want to map \emph{all}\/ the cases
in which there are texture zeros in $M_D$ and $M_R$
while $M_\ell$ remains diagonal
(which in itself means that $M_\ell$ has six texture zeros).
I will look for \emph{predictive}\/ cases,
\textit{i.e.}\ for cases which lead to non-trivial fits
for the lepton mixing (PMNS) matrix and/or for the neutrino mass ratios.
The case in which $M_D$ has six texture zeros
and $M_R$ has two texture zeros was considered in ref.~\cite{Lavoura:2004tu};
here I consider additional cases in which
$M_D$ has either five or four texture zeros and,
correspondingly,
$M_R$ has either three or four texture zeros,
respectively.
In my search I have recovered the cases studied
in refs.~\cite{Dev:2010if} and~\cite{Kageyama:2002zw};
additionally,
I have uncovered some extra cases which had not been,
to my knowledge,
studied before.

Since in my search $M_\ell$ is kept diagonal,
the light neutrinos in the column vector
$\nu = \left( \nu_e,\ \nu_\mu,\ \nu_\tau \right)^T$
may be labelled through their flavour.
Then,
$M= \left[ M_{\alpha \beta} \right]$,
where $\alpha$ and $\beta$ may be either $e$,
$\mu$,
or $\tau$.
In some of the new cases that I present in this paper
the constraints on $M$ may most conveniently be written in terms of the matrix
$A = \left[ A_{\alpha \beta} \right]$ defined by
\be
A_{\alpha \beta} \equiv M_{\alpha \beta} \left( M^{-1} \right)_{\beta \alpha}.
\ee
(I do not use the summation convention in this paper.)
The matrix $A$ was first used in the context of lepton mixing
in refs.~\cite{Ferreira:2013zqa,Ferreira:2013oga}.
It has the properties that the sum of its matrix elements
over any of its rows or columns is equal to one
and that it is invariant either under a rephasing of $M$,
\be
\label{obpdsk}
M_{\alpha \beta} \to e^{i \left( \xi_\alpha + \chi_\beta \right)} M_{\alpha \beta},
\ee
or under the multiplication of $M$
by any number.\footnote{Since $M$ is symmetric,
in its specific case one must set $\chi_\beta = \xi_\beta$ in eq.~(\ref{obpdsk}).}
In our case,
since $M$ is symmetric,
$A$ is symmetric too.
I have found that some texture-zero models predict one diagonal matrix
element of $A$ to be one and another diagonal matrix element of $A$ to be zero;
thus,
but for a permutation of its rows and columns,
\be
\label{a}
A = \left( \begin{array}{ccccc}
0 & & t & & 1-t \\ t & & 1 & & -t \\ 1-t & & -t & & 2t
\end{array} \right),
\ee
where $t$ is some complex number.
In this case,
$A$ has only two degress of freedom
(the real and the imaginary parts of $t$),
instead of the six degrees of freedom of the general case.

\section{Possibilities for $M_D$}

I shall consider
the possibility of a permutation of the rows and columns of $M$
only at the end.\footnote{Both $M$ and $M_R$ are Majorana mass matrices
and therefore they are symmetric.
Hence,
a permutation of their rows
necessarily entails a permutation of their columns too.}
Such a permutation corresponds to a transformation
\be
\label{MZM}
M \to Z M Z^T,
\ee
where
\be
Z \in S_3 \equiv \left\{ A^2,\ A,\ B,\ A B A,\ A B,\ B A \right\},
\ee
where
\be
A \equiv \left( \begin{array}{ccc}
0 & 1 & 0 \\ 1 & 0 & 0 \\ 0 & 0 & 1
\end{array} \right),
\quad
B \equiv \left( \begin{array}{ccc}
0 & 0 & 1 \\ 0 & 1 & 0 \\ 1 & 0 & 0
\end{array} \right).
\ee
The transformation~(\ref{MZM}) is equivalent to
\be
\label{MDZMD}
M_D \to Z M_D,
\ee
which is a permutation of the rows of $M_D$.
Since I am going to consider
the possibility of a transformation~(\ref{MZM}) at the end,
I do not need to consider the possibility of a transformation~(\ref{MDZMD})
now at the beginning.
So,
\emph{I shall take two matrices $M_D$ which only differ
through a permutation of their rows to be equivalent}.

Under this proviso,
a matrix $M_D$ with four texture zeros must be of one of the following forms:
\be
\label{6}
\left( \begin{array}{ccc}
0 & 0 & 0 \\ 0 & \times & \times \\ \times & \times & \times
\end{array} \right),
\quad
\left( \begin{array}{ccc}
0 & 0 & 0 \\ \times & 0 & \times \\ \times & \times & \times
\end{array} \right),
\quad
\left( \begin{array}{ccc}
0 & 0 & 0 \\ \times & \times & 0 \\ \times & \times & \times
\end{array} \right),
\ee
\be
\label{7}
\left( \begin{array}{ccc}
0 & 0 & \times \\ 0 & 0 & \times \\ \times & \times & \times
\end{array} \right),
\quad
\left( \begin{array}{ccc}
0 & \times & 0 \\ 0 & \times & 0 \\ \times & \times & \times
\end{array} \right),
\quad
\left( \begin{array}{ccc}
\times & 0 & 0 \\ \times & 0 & 0 \\ \times & \times & \times
\end{array} \right),
\ee
\be
\label{8}
\left( \begin{array}{ccc}
0 & 0 & \times \\ 0 & \times & 0 \\ \times & \times & \times
\end{array} \right),
\quad
\left( \begin{array}{ccc}
0 & 0 & \times \\ \times & 0 & 0 \\ \times & \times & \times
\end{array} \right),
\quad
\left( \begin{array}{ccc}
0 & \times & 0 \\ \times & 0 & 0 \\ \times & \times & \times
\end{array} \right),
\ee
\be
\label{9}
\begin{array}{c}
{\displaystyle
\left( \begin{array}{ccc}
0 & 0 & \times \\ 0 & \times & \times \\ 0 & \times & \times
\end{array} \right),
\quad
\left( \begin{array}{ccc}
0 & 0 & \times \\ \times & 0 & \times \\ \times & 0 & \times
\end{array} \right),
\quad
\left( \begin{array}{ccc}
0 & \times & 0 \\ 0 & \times & \times \\ 0 & \times & \times
\end{array} \right),
}
\\*[8mm]
{\displaystyle
\left( \begin{array}{ccc}
0 & \times & 0 \\ \times & \times & 0 \\ \times & \times & 0
\end{array} \right),
\quad
\left( \begin{array}{ccc}
\times & 0 & 0 \\ \times & 0 & \times \\ \times & 0 & \times
\end{array} \right),
\quad
\left( \begin{array}{ccc}
\times & 0 & 0 \\ \times & \times & 0 \\ \times & \times & 0
\end{array} \right),
}
\end{array}
\ee
\be
\label{10}
\left( \begin{array}{ccc}
0 & 0 & \times \\ \times & \times & 0 \\ \times & \times & 0
\end{array} \right),
\quad
\left( \begin{array}{ccc}
0 & \times & 0 \\ \times & 0 & \times \\ \times & 0 & \times
\end{array} \right),
\quad
\left( \begin{array}{ccc}
\times & 0 & 0 \\ 0 & \times & \times \\ 0 & \times & \times
\end{array} \right),
\ee
\be
\label{11}
\left( \begin{array}{ccc}
0 & 0 & \times \\ 0 & \times & \times \\ \times & 0 & \times
\end{array} \right),
\quad
\left( \begin{array}{ccc}
0 & \times & 0 \\ 0 & \times & \times \\ \times & \times & 0
\end{array} \right),
\quad
\left( \begin{array}{ccc}
\times & 0 & 0 \\ \times & 0 & \times \\ \times & \times & 0
\end{array} \right),
\ee
\be
\label{11n}
\begin{array}{c}
{\displaystyle
\left( \begin{array}{ccc}
0 & 0 & \times \\ 0 & \times & \times \\ \times & \times & 0
\end{array} \right),
\quad
\left( \begin{array}{ccc}
0 & 0 & \times \\ \times & 0 & \times \\ \times & \times & 0
\end{array} \right),
\quad
\left( \begin{array}{ccc}
0 & \times & 0 \\ 0 & \times & \times \\ \times & 0 & \times
\end{array} \right),
}
\\*[8mm]
{\displaystyle
\left( \begin{array}{ccc}
0 & \times & 0 \\ \times & 0 & \times \\ \times & \times & 0
\end{array} \right),
\quad
\left( \begin{array}{ccc}
\times & 0 & 0 \\ 0 & \times & \times \\ \times & 0 & \times
\end{array} \right),
\quad
\left( \begin{array}{ccc}
\times & 0 & 0 \\ 0 & \times & \times \\ \times & \times & 0
\end{array} \right).
}
\end{array}
\ee
In matrices~(\ref{6}--\ref{11n})
the symbol $\times$ represents a non-zero entry.

If we consider the possibility of transformations~(\ref{MDZMD}),
then there are three matrices of each of the types~(\ref{7}),
(\ref{9}),
and~(\ref{10}),
and six matrices of each of the types~(\ref{6}),
(\ref{8}),
(\ref{11}),
and~(\ref{11n}).
So,
altogether there are $3 \times 12 + 6 \times 15 = 126$
possible matrices $M_D$ with four texture zeros;
this is as it should be,
since $M_D$ has nine independent matrix elements and
$\left( 9 \times 8 \times 7 \times 6 \right) \left/ 4! \right. = 126$.

Matrices $M_D$ with a row of zeros are uninteresting
since they yield one massless,
decoupled neutrino.
Therefore,
the matrices~(\ref{6}) may be neglected.
The matrices~(\ref{7}) may also be neglected because they lead to `scaling',
\textit{i.e.}\ the matrix $M$ has a right-eigenvector with one zero entry
corresponding to the eigenvalue zero~\cite{Mohapatra:2006xy};
this implies that the PMNS matrix has one zero matrix element,
which contradicts experiment.

In order to write down all the possible forms of matrices $M_D$
with five textures zeros,
one simply has to interchange the zeros with the $\times$ symbols
in the matrices~(\ref{6}--\ref{11n}).
One obtains
\be
\label{13}
\left( \begin{array}{ccc}
\times & \times & \times \\ \times & 0 & 0 \\ 0 & 0 & 0
\end{array} \right),
\quad
\left( \begin{array}{ccc}
\times & \times & \times \\ 0 & \times & 0 \\ 0 & 0 & 0
\end{array} \right),
\quad
\left( \begin{array}{ccc}
\times & \times & \times \\ 0 & 0 & \times \\ 0 & 0 & 0
\end{array} \right),
\ee
\be
\label{14}
\left( \begin{array}{ccc}
\times & \times & 0 \\ \times & \times & 0 \\ 0 & 0 & 0
\end{array} \right),
\quad
\left( \begin{array}{ccc}
\times & 0 & \times \\ \times & 0 & \times \\ 0 & 0 & 0
\end{array} \right),
\quad
\left( \begin{array}{ccc}
0 & \times & \times \\ 0 & \times & \times \\ 0 & 0 & 0
\end{array} \right),
\ee
\be
\label{15}
\left( \begin{array}{ccc}
\times & \times & 0 \\ \times & 0 & \times \\ 0 & 0 & 0
\end{array} \right),
\quad
\left( \begin{array}{ccc}
\times & \times & 0 \\ 0 & \times & \times \\ 0 & 0 & 0
\end{array} \right),
\quad
\left( \begin{array}{ccc}
\times & 0 & \times \\ 0 & \times & \times \\ 0 & 0 & 0
\end{array} \right),
\ee
\be
\label{16}
\begin{array}{c}
{\displaystyle
\left( \begin{array}{ccc}
\times & \times & 0 \\ \times & 0 & 0 \\ \times & 0 & 0
\end{array} \right),
\quad
\left( \begin{array}{ccc}
\times & \times & 0 \\ 0 & \times & 0 \\ 0 & \times & 0
\end{array} \right),
\quad
\left( \begin{array}{ccc}
\times & 0 & \times \\ \times & 0 & 0 \\ \times & 0 & 0
\end{array} \right),
}
\\*[8mm]
{\displaystyle
\left( \begin{array}{ccc}
\times & 0 & \times \\ 0 & 0 & \times \\ 0 & 0 & \times
\end{array} \right),
\quad
\left( \begin{array}{ccc}
0 & \times & \times \\ 0 & \times & 0 \\ 0 & \times & 0
\end{array} \right),
\quad
\left( \begin{array}{ccc}
0 & \times & \times \\ 0 & 0 & \times \\ 0 & 0 & \times
\end{array} \right),
}
\end{array}
\ee
\be
\label{17}
\left( \begin{array}{ccc}
\times & \times & 0 \\ 0 & 0 & \times \\ 0 & 0 & \times
\end{array} \right),
\quad
\left( \begin{array}{ccc}
\times & 0 & \times \\ 0 & \times & 0 \\ 0 & \times & 0
\end{array} \right),
\quad
\left( \begin{array}{ccc}
0 & \times & \times \\ \times & 0 & 0 \\ \times & 0 & 0
\end{array} \right),
\ee
\be
\label{18}
\left( \begin{array}{ccc}
\times & \times & 0 \\ \times & 0 & 0 \\ 0 & \times & 0
\end{array} \right),
\quad
\left( \begin{array}{ccc}
\times & 0 & \times \\ \times & 0 & 0 \\ 0 & 0 & \times
\end{array} \right),
\quad
\left( \begin{array}{ccc}
0 & \times & \times \\ 0 & \times & 0 \\ 0 & 0 & \times
\end{array} \right),
\ee
\be
\label{18n}
\begin{array}{c}
{\displaystyle
\left( \begin{array}{ccc}
\times & \times & 0 \\ \times & 0 & 0 \\ 0 & 0 & \times
\end{array} \right),
\quad
\left( \begin{array}{ccc}
\times & \times & 0 \\ 0 & \times & 0 \\ 0 & 0 & \times
\end{array} \right),
\quad
\left( \begin{array}{ccc}
\times & 0 & \times \\ \times & 0 & 0 \\ 0 & \times & 0
\end{array} \right),
}
\\*[8mm]
{\displaystyle
\left( \begin{array}{ccc}
\times & 0 & \times \\ 0 & \times & 0 \\ 0 & 0 & \times
\end{array} \right),
\quad
\left( \begin{array}{ccc}
0 & \times & \times \\ \times & 0 & 0 \\ 0 & \times & 0
\end{array} \right),
\quad
\left( \begin{array}{ccc}
0 & \times & \times \\ \times & 0 & 0 \\ 0 & 0 & \times
\end{array} \right).
}
\end{array}
\ee
Matrices $M_D$ with a row of zeros are uninteresting.
Therefore,
forms~(\ref{13}),
(\ref{14}),
and~(\ref{15}) may be neglected.
Forms~(\ref{16}) and~(\ref{17}) may also be neglected
because they lead to scaling.
Therefore,
only the nine forms~(\ref{18}) and~(\ref{18n}) should be considered.

\section{Possibilities for $M_R$}

The matrix $M$ is invariant under
\be
M_D \to M_D Z,
\quad
M_R \to Z^T\! M_R Z
\ee
because $Z^T = Z^{-1},\ \forall Z \in S_3$.
Therefore,
a permutation of the rows and columns of $M_R$
is equivalent to a permutation of the columns of $M_D$.
Since in the preceding section
I have not chosen any particular order for the columns of $M_D$,
I am free in this section
to restrict the order of the rows and columns of $M_R$.
Moreover,
if one particular form of $M_R$ is invariant under some $S_2$ subgroup of $S_3$,
then one may disconsider the action of that $S_2$ on the columns of $M_D$.

If one wants to obtain a model as predictive
as those in the literature\footnote{The models in ref.~\cite{Lavoura:2004tu}
have six zeros in $M_D$ and two zeros in $M_R$.
Models as predictive should have either
five zeros in $M_D$ and three zeros in $M_R$ or
four zeros in $M_D$ and four zeros in $M_R$.}
and if $M_D$ has four texture zeros,
then $M_R$ should also have four texture zeros.
Since $\det{M_R}$ must be nonzero,\footnote{If $M_R$ is a singular matrix
then one right-handed neutrino is massless and the see-saw mechanism
is not fully operative; see ref.~\cite{Branco:1988ex}.}
there is only one possible form for $M_R$,
\be
\label{MR}
\left( \begin{array}{ccc}
\times & 0 & 0 \\ 0 & 0 & \times \\ 0 & \times & 0
\end{array} \right),
\ee
but for permutations of the rows and columns---which are equivalent to
permutations of the columns of $M_D$.
Equation~(\ref{MR}) leads to
\be
\label{igopr}
M_R^{-1} = \left( \begin{array}{ccc}
x & 0 & 0 \\ 0 & 0 & y \\ 0 & y & 0
\end{array} \right).
\ee
The form~(\ref{MR}) of $M_R$ is invariant under the interchange
of the second and third rows and columns.
Therefore,
when $M_R$ is of that form,
one may disconsider
the possibility of a permutation of the second and third columns of $M_D$.
Thus,
out of the 21 forms~(\ref{8}--\ref{11n}) of $M_D$
only the following 13 must be considered:
\bs
\label{24}
\ba
\label{24a}
M_D &=& \left( \begin{array}{ccc}
0 & 0 & a \\ 0 & b & 0 \\ c & d & e
\end{array} \right),
\\
\label{24b}
M_D &=& \left( \begin{array}{ccc}
0 & 0 & a \\ b & 0 & 0 \\ c & d & e
\end{array} \right);
\ea
\es
\bs
\label{25}
\ba
\label{25a}
M_D &=& \left( \begin{array}{ccc}
0 & 0 & a \\ 0 & b & c \\ 0 & d & e
\end{array} \right),
\\
\label{25b}
M_D &=& \left( \begin{array}{ccc}
0 & 0 & a \\ b & 0 & c \\ d & 0 & e
\end{array} \right),
\\
\label{25c}
M_D &=& \left( \begin{array}{ccc}
a & 0 & 0 \\ b & 0 & c \\ d & 0 & e
\end{array} \right);
\ea
\es
\bs
\label{26}
\ba
\label{26a}
M_D &=& \left( \begin{array}{ccc}
0 & 0 & a \\ b & c & 0 \\ d & e & 0
\end{array} \right),
\\
\label{26b}
M_D &=& \left( \begin{array}{ccc}
a & 0 & 0 \\ 0 & b & c \\ 0 & d & e
\end{array} \right);
\ea
\es
\bs
\label{27}
\ba
\label{27a}
M_D &=& \left( \begin{array}{ccc}
0 & 0 & a \\ 0 & b & c \\ d & 0 & e
\end{array} \right),
\\
\label{27b}
M_D &=& \left( \begin{array}{ccc}
0 & 0 & a \\ 0 & b & c \\ d & e & 0
\end{array} \right),
\\
\label{27c}
M_D &=& \left( \begin{array}{ccc}
0 & 0 & a \\ b & 0 & c \\ d & e & 0
\end{array} \right),
\\
\label{27d}
M_D &=& \left( \begin{array}{ccc}
0 & a & 0 \\ b & 0 & c \\ d & e & 0
\end{array} \right),
\\
\label{27e}
M_D &=& \left( \begin{array}{ccc}
a & 0 & 0 \\ 0 & b & c \\ d & 0 & e
\end{array} \right),
\\
\label{27f}
M_D &=& \left( \begin{array}{ccc}
a & 0 & 0 \\ b & 0 & c \\ d & e & 0
\end{array} \right).
\ea
\es

If $M_D$ has five texture zeros then $M_R$ should have three texture zeros.
Since $\det{M_R}$ must be nonzero,
the following are the only possible forms for $M_R$:
\bs
\ba
\label{u2}
\left( \begin{array}{ccc}
\times & 0 & 0 \\ 0 & \times & 0 \\ 0 & 0 & \times
\end{array} \right),
& &
\left( \begin{array}{ccc}
0 & \times & \times \\ \times & 0 & \times \\ \times & \times & 0
\end{array} \right),
\\
\label{u3}
\left( \begin{array}{ccc}
\times & \times & 0 \\ \times & 0 & 0 \\ 0 & 0 & \times
\end{array} \right),
& & \left( \begin{array}{ccc}
\times & \times & 0 \\ \times & 0 & \times \\ 0 & \times & 0
\end{array} \right).
\ea
\es
They correspond to
\bs
\ba
\label{f1}
M_R^{-1} &=& \left( \begin{array}{ccc}
x & 0 & 0 \\ 0 & y & 0 \\ 0 & 0 & z
\end{array} \right),
\\
\label{f4}
M_R^{-1} &=& \left( \begin{array}{ccc}
- x & \sqrt{x y} & \sqrt{x z} \\
\sqrt{x y} & - y & \sqrt{y z} \\
\sqrt{x z} & \sqrt{y z} & - z
\end{array} \right),
\\
\label{f2}
M_R^{-1} &=& \left( \begin{array}{ccc}
0 & x & 0 \\ x & y & 0 \\ 0 & 0 & z
\end{array} \right),
\\
\label{f3}
M_R^{-1} &=& \left( \begin{array}{ccc}
x & 0 & \sqrt{x z} \\ 0 & 0 & y \\ \sqrt{x z} & y & z
\end{array} \right),
\ea
\es
respectively.

The forms of the matrices~(\ref{u2}) are invariant under $S_3$
while the forms of the matrices~(\ref{u3}) are \emph{not}\/ invariant
under any non-trivial permutation of their rows and columns.
Therefore,
with either eq.~(\ref{f1}) or eq.~(\ref{f4}) one may take,
instead of the nine possibilities~(\ref{18}--\ref{18n}) for $M_D$,
just the following two possibilities:
\bs
\label{ibvuty}
\ba
\label{ibvuty3}
M_D &=& \left( \begin{array}{ccc}
a & b & 0 \\ c & 0 & 0 \\ 0 & d & 0
\end{array} \right),
\\
\label{ibvuty4}
M_D &=& \left( \begin{array}{ccc}
a & b & 0 \\ c & 0 & 0 \\ 0 & 0 & d
\end{array} \right).
\ea
\es
With eqs.~(\ref{f2}) and~(\ref{f3}),
on the other hand,
one must use the full set of nine possibilities for $M_D$:
\bs
\label{iby}
\ba
\label{iby1}
M_D &=& \left( \begin{array}{ccc}
a & b & 0 \\ c & 0 & 0 \\ 0 & d & 0
\end{array} \right),
\\
\label{iby2}
M_D &=& \left( \begin{array}{ccc}
a & 0 & b \\ c & 0 & 0 \\ 0 & 0 & d
\end{array} \right),
\\
\label{iby3}
M_D &=& \left( \begin{array}{ccc}
0 & a & b \\ 0 & c & 0 \\ 0 & 0 & d
\end{array} \right),
\ea
\es
\bs
\label{ibi}
\ba
\label{ibi1}
M_D &=& \left( \begin{array}{ccc}
a & b & 0 \\ c & 0 & 0 \\ 0 & 0 & d
\end{array} \right),
\\
\label{ibi2}
M_D &=& \left( \begin{array}{ccc}
a & b & 0 \\ 0 & c & 0 \\ 0 & 0 & d
\end{array} \right),
\\
\label{ibi3}
M_D &=& \left( \begin{array}{ccc}
a & 0 & b \\ c & 0 & 0 \\ 0 & d & 0
\end{array} \right),
\\
\label{ibi4}
M_D &=& \left( \begin{array}{ccc}
a & 0 & b \\ 0 & c & 0 \\ 0 & 0 & d
\end{array} \right),
\\
\label{ibi5}
M_D &=& \left( \begin{array}{ccc}
0 & a & b \\ c & 0 & 0 \\ 0 & d & 0
\end{array} \right),
\\
\label{ibi6}
M_D &=& \left( \begin{array}{ccc}
0 & a & b \\ c & 0 & 0 \\ 0 & 0 & d
\end{array} \right).
\ea
\es

\section{Constraints on $M$}

\subsection{Possibilities with eq.~(\ref{f1})}

With this form of $M_R^{-1}$ one should use the two
options~(\ref{ibvuty}) for $M_D$.
With eq.~(\ref{ibvuty4}) one neutrino decouples;
this is incompatible with experiment.
With eq.~(\ref{ibvuty3}) one obtains\footnote{To be sure,
from eqs.~(\ref{f1}) and~(\ref{ibvuty3})
one obtains $\det{M} = 0$ and $M_{23} = 0$.
But now I generalize and consider other options for $M_D$
that differ from eq.~(\ref{ibvuty3}) through a permutation of the rows;
then I obtain,
in general,
the conditions~(\ref{uit}).
In the same fashion,
throughout this section I shall consider,
for each particular case,
the results that follow after considering all possible permutations
of the rows of $M_D$.}
\be
\label{uit}
\det{M} = 0, \quad M_{\alpha \beta} = 0\ (\alpha \neq \beta),
\ee
which are constraints on $M$ which have not yet
been considered in the literature.

\subsection{Possibilities with eq.~(\ref{f4})}

With this form of $M_R^{-1}$ one should once again use
the two options~(\ref{ibvuty}) for $M_D$.
With eq.~(\ref{ibvuty4}) one obtains
\be
\label{obpt}
\left( M^{-1} \right)_{\alpha \alpha} = \left( M^{-1} \right)_{\beta \beta} = 0
\quad (\alpha \neq \beta),
\ee
which has already been considered in ref.~\cite{Lavoura:2004tu}.
With eq.~(\ref{ibvuty3}) one obtains
\be
\det{M} = 0, \quad
M_{\alpha \alpha} M_{\beta \beta} - \left( M_{\alpha \beta} \right)^2 = 0
\quad (\alpha \neq \beta),
\ee
which is new and potentially interesting.

\subsection{Possibilities with eq.~(\ref{f2})}

With this form of $M_R^{-1}$ one should use
either one of the nine options~(\ref{iby}--\ref{ibi}) for $M_D$.

With eq.~(\ref{iby2}) one of the neutrinos decouples.
With eq.~(\ref{iby3}) one recovers the conditions~(\ref{uit}).
With eq.~(\ref{iby1}) one obtains
\be
\label{uit2}
\det{M} = 0, \quad M_{\alpha \alpha} = 0,
\ee
which must be studied.

Both eq.~(\ref{ibi1}) and eq.~(\ref{ibi2})
lead to the decoupling of one neutrino.
With either eq.~(\ref{ibi3}) or eq.~(\ref{ibi6}) one gets
\be
\label{svort}
M_{\alpha \alpha} = M_{\alpha \beta} = 0
\quad (\alpha \neq \beta),
\ee
which has been studied in ref.~\cite{Frampton:2002yf}.
With eq.~(\ref{ibi4}) one has
\be
\label{nutr1}
\left( M^{-1} \right)_{\alpha \alpha} = 0, \quad
M_{\alpha \beta} = 0 \quad (\alpha \neq \beta),
\ee
while with eq.~(\ref{ibi5}) one has
\be
\label{nutr2}
M_{\alpha \alpha} = 0, \quad
\left( M^{-1} \right)_{\alpha \beta} = 0 \quad (\alpha \neq \beta).
\ee
Both constraints~(\ref{nutr1}) and~(\ref{nutr2})
have already been studied in ref.~\cite{Dev:2010if}.

\subsection{Possibilities with eq.~(\ref{f3})}

With this form of $M_R^{-1}$ one should use
either one of the nine options~(\ref{iby}--\ref{ibi}) for $M_D$.

Equation~(\ref{iby1}) leads to  one neutrino decoupling
and eq.~(\ref{iby2}) leads to scaling.
Equation~(\ref{iby3}) reproduces the conditions~(\ref{uit2}).

Equation~(\ref{ibi1}) leads once again to the conditions~(\ref{obpt}).
Both eq.~(\ref{ibi2}) and eq.~(\ref{ibi5})
reproduce the conditions~(\ref{svort}).
Equation~(\ref{ibi4}) leads to
\be
\label{biuft}
M_{\alpha \alpha} = 0, \quad  \left( M^{-1} \right)_{\alpha \alpha} = 0.
\ee
These conditions have also been studied in ref~\cite{Dev:2010if}.
Equation~(\ref{ibi3}) leads to
\be
M_{\alpha \alpha} = M_{\alpha \beta} = 0, \quad 
\left( M^{-1} \right)_{\alpha \alpha} = 0 \quad (\alpha \neq \beta).
\ee
This is one constraint too many,
but I shall consider it later.
Equation~(\ref{ibi6}) gives
\be
\label{blpde}
\left( M^{-1} \right)_{\alpha \alpha} = 0,
\quad
A_{\beta \beta} = 1,
\quad
M_{\gamma \gamma} \neq 0
\quad
(\alpha \neq \beta \neq \gamma \neq \alpha),
\ee
which is new.
I have explicitly written down
the condition $M_{\gamma \gamma} \neq 0$ in conditions~(\ref{blpde})
in order to distinguish this model from cases $A_{1,2}$ and $B_{3,4}$
of ref.~\cite{Frampton:2002yf}.
For instance,
$M_{11} = M_{12} = 0$ in case $A_1$;
this leads to $\left( M^{-1} \right)_{33} = 0$ and $A_{22} = 1$,
corresponding to $t = 0$ in eq.~(\ref{a}).

\subsection{Possibilities with eq.~(\ref{igopr})}

With eq.~(\ref{igopr}) one must use for $M_D$ the 13 options
in eqs.~(\ref{24}--\ref{27}).

Equation~(\ref{24a}) yields
\be
M_{\alpha \alpha} = M_{\beta \beta} = 0
\quad (\alpha \neq \beta).
\ee
These conditions have been studied in ref.~\cite{Frampton:2002yf}
(see also ref.~\cite{Grimus:2004az}).
Equation~(\ref{24b}) reproduces the conditions~(\ref{svort}).

Equation~(\ref{25a}) yields once again the conditions~(\ref{uit2}).
Equations~(\ref{25b}) and~(\ref{25c}) lead to two massless neutrinos.

Equation~(\ref{26a}) reproduces the conditions~(\ref{biuft}).
Equation~(\ref{26b}) makes one neutrino decouple.

Equations~(\ref{27a}),
(\ref{27c}),
and~(\ref{27d}) reproduce the conditions~(\ref{svort}).
Equation~(\ref{27f}) reproduces the conditions~(\ref{obpt}).
Equation~(\ref{27e}) reproduces the conditions~(\ref{nutr1}).
Finally,
eq.~(\ref{27b}) yields
\be
\label{kpft}
M_{\alpha \alpha} = 0, \quad A_{\beta \beta} = 1,
\quad
\left( M^{-1} \right)_{\gamma \gamma} \neq 0
\quad
(\alpha \neq \beta \neq \gamma \neq \alpha),
\ee
which is new.

\subsection{Summary}

Most matrices $M$ that I have found embody conditions
that have already been treated in the literature.
A few matrices $M$,
though,
present features which,
to my knowledge,
have not yet been studied.
These are
\ba
\label{o1}
& & \det{M} = 0 \quad \mbox{and} \quad M_{\alpha \alpha} = 0;
\\
\label{o2}
& & \det{M} = 0 \quad \mbox{and} \quad M_{\alpha \beta} = 0;
\\
\label{o3}
& & \det{M} = 0 \quad \mbox{and} \quad
M_{\alpha \alpha} M_{\beta \beta} - \left( M_{\alpha \beta} \right)^2 = 0;
\\
\label{extra}
& & M_{\alpha \alpha} = M_{\alpha \beta} = 0 \quad \mbox{and} \quad
\left( M^{-1} \right)_{\alpha \alpha} = 0,
\\
\label{o4}
& & M_{\alpha \alpha} = 0, \quad A_{\beta \beta} = 1, \quad \mbox{and} \quad
\left( M^{-1} \right)_{\gamma \gamma} \neq 0;
\\
\label{o5}
& & \left( M^{-1} \right)_{\alpha \alpha} = 0, \quad A_{\beta \beta} = 1,
\quad \mbox{and} \quad M_{\gamma \gamma} \neq 0.
\ea
In eqs.~(\ref{o2}--\ref{o5}) it should be understood that
$\alpha \neq \beta \neq \gamma \neq \alpha$.

According to ref.~\cite{Dev:2010if},
the possibility~(\ref{extra}) should be excluded because
$M_{\alpha \alpha} = 0$ together with $\left( M^{-1} \right)_{\alpha \alpha} = 0$
is experimentally excluded for any value of $\alpha = e, \mu, \tau$.
So in the next section I shall only consider conditions~(\ref{o1}--\ref{o3}),
(\ref{o4}),
and~(\ref{o5}).

\section{Comparison with the data}

\subsection{Introduction}

\paragraph{PMNS matrix:}
Since $M_\ell$ is diagonal,
the unitary matrix that diagonalizes $M$ is the lepton mixing (PMNS) matrix $U$:
\be
M = U^\ast\, \mbox{diag} \left( \mu_1,\ \mu_2,\ \mu_3 \right) U^\dagger,
\ee
where the $\mu_j$ are complex;
the neutrino masses are $ m_j = \left| \mu_j \right|$
($j = 1, 2, 3$).
The matrix $U$ is written
\be
\label{U}
U = \left( \begin{array}{ccc}
c_{12} c_{13} & s_{12} c_{13} & \epsilon^\ast \\
- s_{12} c_{23} - \epsilon c_{12} s_{23} &
c_{12} c_{23} - \epsilon s_{12} s_{23} &
s_{23} c_{13} \\
s_{12} s_{23} - \epsilon c_{12} c_{23} &
- c_{12} s_{23} - \epsilon s_{12} c_{23} &
c_{23} c_{13}
\end{array} \right),
\ee
where $\epsilon \equiv s_{13} \exp{\left( i \delta \right)}$.
In eq.~(\ref{U}),
$s_{jj^\prime} \equiv \sin{\theta_{jj^\prime}}$
and $c_{jj^\prime} \equiv \cos{\theta_{jj^\prime}}$.

\paragraph{The data:}
I define
\be
r_\mathrm{solar} \equiv \sqrt{\frac{m_2^2 - m_1^2}{\left| m_3^2 - m_1^2 \right|}}.
\ee
I use the $3 \sigma$ ranges~\cite{Forero:2014bxa}\footnote{There are other
phenomenological fits to the data---see
refs.~\cite{Fogli:2012ua,Gonzalez-Garcia:2014bfa}.}
\be
\label{normalranges}
\begin{array}{rcccl}
0.278 & \le & s_{12}^2 & \le & 0.375, \\*[1mm]
0.0177 & \le & s_{13}^2 & \le & 0.0294, \\*[1mm]
0.392 & \le & s_{23}^2 & \le & 0.643, \\*[1mm]
0.0268 & \le & r_\mathrm{solar}^2 & \le & 0.0356
\end{array}
\ee
for `normal' ordering of the neutrino masses ($m_3 > m_2 > m_1$),
and
\be
\label{invertedranges}
\begin{array}{rcccl}
0.278 & \le & s_{12}^2 & \le & 0.375, \\*[1mm]
0.0183 & \le & s_{13}^2 & \le & 0.0297, \\*[1mm]
0.403 & \le & s_{23}^2 & \le & 0.640, \\*[1mm]
0.0280 & \le & r_\mathrm{solar}^2 & \le & 0.0372
\end{array}
\ee
for `inverted' ordering ($m_2 > m_1 > m_3$).

\paragraph{Neutrino mass observables:}
The models in this paper
cannot predict the absolute value of the neutrino masses,
since all the predictions in eqs.~(\ref{o1})--(\ref{o5})
are invariant under $M \to c M$,
where $c$ is an arbitrary complex number.
They may,
though,
predict the \emph{relative}\/ value of any two neutrino mass observables.
Most conveniently,
one of those observables should be chosen to be
the square root of the atmospheric squared-mass difference,
\be
m_\mathrm{atmospheric} \equiv \sqrt{\left| m_3^2 - m_1^2 \right|}
\approx 0.05\, \mathrm{eV}.
\ee
The other relevant mass observables---besides
$m_\mathrm{solar} \equiv \sqrt{m_2^2 - m_1^2}$---are
\bs
\ba
m_\mathrm{cosmological} &\equiv& m_1 + m_2 + m_3, \\
m_{\beta \beta} &\equiv& \left| M_{ee} \right|
= \left| \sum_{j=1}^3 \mu_j^\ast \left( U_{ej} \right)^2 \right|, \\
m_{\nu_e} &\equiv& \sum_{j=1}^3 m_j \left| U_{ej} \right|^2.
\ea
\es
Indeed,
$m_\mathrm{cosmological}$ may be derived from various cosmological observations;
$m_{\beta \beta}$ may be derived
from the rates of neutrinoless double-$\beta$ decay of various nuclides;
and $\left\langle m_{\nu_e} \right\rangle$
is the average mass of the electron neutrino
to be measured in experiments on the electron energy end-point
of tritium $\beta$ decay.
I define
\be
r_\mathrm{cosmological} \equiv \frac{m_\mathrm{cosmological}}
{m_\mathrm{atmospheric}},
\quad
r_{\beta \beta} \equiv \frac{m_{\beta \beta}}{m_\mathrm{atmospheric}},
\quad
r_{\nu_e} \equiv \frac{m_{\nu_e}}{m_\mathrm{atmospheric}}.
\ee

\subsection{The conditions~(\ref{o3})}

When $\det{M} = 0$ either $\mu_1 = 0$ (normal ordering)
or $\mu_3 = 0$ (inverted ordering).
With normal ordering one has
\ba
0 &=& M^\ast_{\alpha \alpha} M^\ast_{\beta \beta} - \left( M^\ast_{\alpha \beta} \right)^2
\no &=&
\left[ \mu_2^\ast \left( U_{\alpha 2} \right)^2
+ \mu_3^\ast \left( U_{\alpha 3} \right)^2 \right]
\left[ \mu_2^\ast \left( U_{\beta 2} \right)^2
+ \mu_3^\ast \left( U_{\beta 3} \right)^2 \right]
\no & &
- \left( \mu_2^\ast U_{\alpha 2} U_{\beta 2}
+ \mu_3^\ast U_{\alpha 3} U_{\beta 3} \right)^2.
\ea
This gives
\be
0 = \mu_2^\ast \mu_3^\ast
\left( U_{\alpha 2} U_{\beta 3} - U_{\alpha 3} U_{\beta 2} \right)^2,
\ee
hence $U_{\alpha 2} U_{\beta 3} - U_{\alpha 3} U_{\beta 2} = 0$.
But $U$ is a unitary matrix,
therefore
\be
\left| U_{\alpha 2} U_{\beta 3} - U_{\alpha 3} U_{\beta 2} \right|
= \left| U_{\gamma 1} \right|,
\ee
where $\gamma \neq \alpha, \beta$.
One concludes that the conditions~(\ref{o3}) predict,
in the case of normal ordering,
one matrix element of the first column of $U$ to vanish.
This contradicts experiment.
In the case of inverse ordering,
conditions~(\ref{o3}) predict a matrix element of the third column of $U$
to vanish.
This also contradicts experiment.

Thus,
conditions~(\ref{o3}) are experimentally excluded.

\subsection{The conditions~(\ref{o1})}

With normal ordering one has $\mu_1 = 0$ and
\be
0 = M^\ast_{\alpha \alpha} = \mu_2^\ast \left( U_{\alpha 2} \right)^2
+ \mu_3^\ast \left( U_{\alpha 3} \right)^2.
\ee
Therefore,
\be
\label{cond1}
\left| \frac{U_{\alpha 3}}{U_{\alpha 2}} \right|^2
= \frac{m_2}{m_3} = r_\mathrm{solar}.
\ee
Equation~(\ref{cond1}) is incompatible with experiment
for any $\alpha = e, \mu, \tau$.

With inverted ordering one has instead $\mu_3 = 0$ and
\be
0 = M^\ast_{\alpha \alpha} = \mu_1^\ast \left( U_{\alpha 1} \right)^2
+ \mu_2^\ast \left( U_{\alpha 2} \right)^2.
\ee
Therefore,
\be
\label{cond2}
\left| \frac{U_{\alpha 1}}{U_{\alpha 2}} \right|^2 = \frac{m_2}{m_1}
= \sqrt{1 + r_\mathrm{solar}^2}.
\ee
Equation~(\ref{cond2}) can fit the phenomenology
in the cases $\alpha = \mu$ and $\alpha = \tau$.
If $\alpha = \mu$,
then $s_{12}$ should not be too low;
$s_{23}$ and (to a lesser extent) $s_{13}$ are also preferably
above their central values;
moreover,
$\cos{\delta} \gtrsim 0.5$ is predicted.
If $\alpha = \tau$,
then $s_{12}$ and $s_{13}$ should be at or above their best-fit values
while $\theta_{23}$ lies preferably in the first octant;
$\cos{\delta} \lesssim -0.5$ in this case.
In both cases,
the predictions for the neutrino mass ratios are
\be
\label{pred23}
2.014 \le r_\mathrm{cosmological} \le 2.018, \quad
0.24 \le r_{\beta \beta} \le 0.42, \quad
0.974 \le r_{\nu_e} \le 0.988.
\ee

To summarize,
conditions~(\ref{o1}) can only hold with an inverted neutrino mass spectrum,
with either $\alpha = \mu$ or $\alpha = \tau$,
and with large $\left| \cos{\delta} \right|$.

\subsection{The conditions~(\ref{o2})}

With normal ordering the conditions~(\ref{o2}) produce
\be
\label{cond3}
\left| \frac{U_{\alpha 3} U_{\beta 3}}{U_{\alpha 2} U_{\beta 2}} \right|
= r_\mathrm{solar}
\ee
while with inverted ordering one obtains instead
\be
\label{cond4}
\left| \frac{U_{\alpha 1} U_{\beta 1}}{U_{\alpha 2} U_{\beta 2}} \right|
= \sqrt{1 + r_\mathrm{solar}^2}.
\ee
Equation~(\ref{cond3}) is incompatible with experiment.
Condition~(\ref{cond4}) may agree with the phenomenology
in the cases $\alpha = e$ and either $\beta = \mu$ or $\beta = \tau$;
the mixing angles are free but there is a stringent prediction
$\left| \cos{\delta} \right| < 0.1$.
The predictions for $r_\mathrm{cosmological}$ and $r_{\nu_e}$
are the same as in inequalities~(\ref{pred23})
while $0.948 \le r_{\beta \beta} \le 0.983$ is higher in this case.

To summarize,
conditions~(\ref{o2}) can only hold if the neutrino mass spectrum is inverted
and if $\left( \alpha, \beta \right)$
is either $\left( e, \mu \right)$ or $\left( e, \tau \right)$.
A tiny $\left| \cos{\delta} \right|$ is predicted.

\subsection{The conditions~(\ref{o4})}

I firstly define
\be
\label{xy}
V_{\beta j} \equiv \left( U_{\beta j}^\ast \right)^2,
\quad
x \equiv \frac{\mu_1}{\mu_3},
\quad
y \equiv \frac{\mu_2}{\mu_3}.
\ee
I then define
\bs
\ba
c_1 &\equiv& V_{\alpha 1}, \\
c_2 &\equiv& V_{\alpha 2}, \\
c_3 &\equiv& V_{\alpha 3}, \\
c_4 &\equiv& V_{\beta 1} V_{\beta 2}^\ast, \quad c_5 \equiv c_4^\ast, \\
c_6 &\equiv& V_{\beta 1} V_{\beta 3}^\ast, \quad c_7 \equiv c_6^\ast, \\
c_8 &\equiv& V_{\beta 2} V_{\beta 3}^\ast, \quad c_9 \equiv c_8^\ast,\\
c_{10} &\equiv& \sum_{j=1}^3 \left| V_{\beta j} \right|^2 - 1.
\ea
\es
Then,
from the first condition~(\ref{o4}),
\be
0 = M_{\alpha \alpha} = \mu_1 V_{\alpha 1} + \mu_2 V_{\alpha 2} + \mu_3 V_{\alpha 3}.
\ee
Therefore,
\be
\label{iorex}
0 = c_1 x + c_2 y + c_3.
\ee
The second condition~(\ref{o4}) is
\ba
1 &=& A_{\beta \beta}
\no &=& M_{\beta \beta} \left( M^{-1} \right)_{\beta \beta}
\no &=& \left(
\mu_1 V_{\beta 1} + \mu_2 V_{\beta 2} + \mu_3 V_{\beta 3}
\right) \left(
\frac{V_{\beta 1}^\ast}{\mu_1} + \frac{V_{\beta 2}^\ast}{\mu_2}
+ \frac{V_{\beta 3}^\ast}{\mu_3}
\right)
\no &=& \left| V_{\beta 1} \right|^2 + \left| V_{\beta 2} \right|^2
+ \left| V_{\beta 3} \right|^2
+ \frac{y}{x}\, V_{\beta 1}^\ast V_{\beta 2}
+ \frac{x}{y}\, V_{\beta 1} V_{\beta 2}^\ast
\no & &
+ \frac{1}{x}\, V_{\beta 1}^\ast V_{\beta 3}
+ x V_{\beta 1} V_{\beta 3}^\ast
+ \frac{1}{y}\, V_{\beta 2}^\ast V_{\beta 3}
+ y V_{\beta 2} V_{\beta 3}^\ast.
\ea
Therefore,
\be
0 = c_4 x^2 + c_5 y^2 + c_6 x^2 y + c_7 y + c_8 x y^2 + c_9 x + c_{10} x y.
\label{uviop}
\ee
Equations~(\ref{iorex}) and~(\ref{uviop})
determine $x$ and $y$ through
\bs
\label{jo}
\ba
0 &=& c_2 \left( c_2 c_6 - c_1 c_8 \right) y^3
\no & &
+ \left( 2 c_2 c_3 c_6 - c_1 c_2 c_{10} - c_1 c_3 c_8 + c_1^2 c_5
+ c_2^2 c_4 \right) y^2
\no & &
+ \left( 2 c_2 c_3 c_4 - c_1 c_3 c_{10} - c_1 c_2 c_9 + c_1^2 c_7
+ c_3^2 c_6 \right) y
\no & &
+ c_3 \left( c_3 c_4 - c_1 c_9 \right),
\\*[1mm]
x &=& \frac{- c_2 y - c_3}{c_1}.
\ea
\es
In this way,
the first two conditions~(\ref{o4}) allow one to,
by using as input the PMNS matrix,
exactly determine both the Majorana phases
and the ratios among the neutrino masses.
One must still impose the third condition~(\ref{o4}),
\textit{viz.}
\be
V_{\gamma 1}^\ast y + V_{\gamma 2}^\ast x + V_{\gamma 3}^\ast x y \neq 0,
\ee
on the values of $x$ and $y$ that have been determined.

One must choose the input,
\textit{viz.}\ the PMNS matrix,
in such a way that the resulting $x$ and $y$ satisfy
$\left| x \right| < \left| y \right|$,
\textit{i.e.}\ $m_\mathrm{solar} > 0$,
and that
\be
r_\mathrm{solar} = \sqrt{\frac{\left| y \right|^2 - \left| x \right|^2}
{\left| 1 - \left| x \right|^2 \right|}}
\ee
is in its experimentally allowed range.
If $1 > \left| x \right|$ the neutrino mass spectrum is normal;
it is inverted if $\left| x \right| > 1$.
For the neutrino mass ratios one has
\bs
\ba
r_\mathrm{cosmological} &=&
\frac{\left| x \right| + \left| y \right| + 1}
{\sqrt{\left| 1 - \left| x \right|^2 \right|}},
\\
r_{\beta \beta} &=&
\frac{\left| x V_{e1} + y V_{e2} + V_{e3} \right|}
{\sqrt{\left| 1 - \left| x \right|^2 \right|}},
\\
r_{\nu_e} &=&
\frac{\left| x V_{e1} \right| + \left| y V_{e2} \right|
+ \left| V_{e3} \right|}
{\sqrt{\left| 1 - \left| x \right|^2 \right|}}.
\ea
\es

Numerically,
I have found that there are two types of cases
in which the conditions~(\ref{o4}) are able to fit the experimental values.
In the first type of cases,
$\left| \cos{\delta} \right| \gtrsim 0.5$ must be close
to unity\footnote{In this paper,
the expression ``$a \gtrsim b$'' means the following:
the quantity $a$ has a lower bound that is approximately equal to $b$,
but $a$ may as well be much larger than $b$.} 
and the neutrino masses are of order $m_\mathrm{atmospheric}$.
This type of cases occurs for an inverted neutrino mass spectrum
when $\beta = e$ and either $\alpha = \mu$ or $\alpha = \tau$;
in the first case $\cos{\delta} \gtrsim 0.5$
and in the second one $\cos{\delta} \lesssim -0.5$.
One obtains for these cases
\be
\begin{array}{rcccl}
2.019 &<& r_\mathrm{cosmological} &<& 2.035, \\
0.24 &<& r_{\beta \beta} &<& 0.44, \\
0.974 &<& r_{\nu_e} &<& 0.989.
\end{array}
\ee

In the second type of cases neutrino masses are quasi-degenerate,
$\theta_{23}$ is in a well-defined octant,
and $\cos{\delta}$ is extremely close to zero.
Moreover,
when $\theta_{23} \to \pi/4$,
the neutrino masses grow towards infinity
and $\cos{\delta} \to 0$.\footnote{In ref.~\cite{Grimus:2011sf}
it had already been noted that models $B_{3,4}$ of ref.~\cite{Frampton:2002yf}
display the property that $\theta_{2,3} \to \pi/4$ and $\cos{\delta} \to 0$
when the neutrino masses become quasi-degenerate.
Our models,
though,
do \emph{not}\/ coincide with models $B_{3,4}$.
In those models $A_{ee} = 1$ and $A_{\mu \mu} = A_{\tau \tau} = 0$;
in our models $A_{ee} = 1$ and either $A_{\mu \mu}$ vanishes
or $A_{\tau \tau}$ vanishes,
but they do not \emph{both}\/ vanish.}
These cases occur when $\beta = e$ and
either $\alpha = \mu$ for a normal neutrino mass spectrum
or $\alpha = \tau$ for an inverted neutrino mass spectrum;
in both cases $s_{23}^2 < 0.5$.
If one wants to have $s_{23}^2 > 0.5$ instead,
then we must interchange $\alpha = \mu$ with $\alpha = \tau$.
In all these cases $\left| \cos{\delta} \right| < 0.1$
(approaching zero when $s_{23}^2$ approaches $0.5$)
and $r_\mathrm{cosmological} \gtrsim 2.5$,
$r_{\beta \beta} \approx r_{\nu_e} \gtrsim 0.5$
(approaching infinity when $s_{23}^2$ approaches $0.5$).

\subsection{The conditions~(\ref{o5})}

Equations~(\ref{o5}) are the same as eqs.~(\ref{o4})
with $M \leftrightarrow M^{-1}$,
\textit{i.e.}\ with $\mu_j \to \mu_j^{-1}$ and $V \to V^\ast$.
In practice,
this means that,
for each input PMNS matrix,
one may use eqs.~(\ref{jo}) in the previous subsection,
but now they will yield $1 \! \left/ x^\ast \right.$
and $1 \! \left/ y^\ast \right.$ instead of $x$ and $y$,
respectively.
Thus,
one should use
\be
V_{\beta j} \equiv \left( U_{\beta j}^\ast \right)^2,
\quad
x \equiv \frac{\mu_3^\ast}{\mu_1^\ast},
\quad
y \equiv \frac{\mu_3^\ast}{\mu_2^\ast}
\ee
instead of eqs.~(\ref{xy}).
The condition $m_\mathrm{solar} > 0$
now requires $\left| x \right| > \left| y \right|$
and
\be
r_\mathrm{solar} = \sqrt{\frac{\left| x \right|^2 - \left| y \right|^2}
{\left| y \right|^2 \left| \left| x \right|^2 - 1 \right|}}
\ee
must be in its experimentally allowed range.
If $1 > \left| x \right|$ then the neutrino mass spectrum is inverted;
it is normal if $\left| x \right| > 1$.
For the neutrino mass ratios one has in this case
\bs
\ba
r_\mathrm{cosmological} &=&
\frac{\left| x \right| + \left| y \right| + \left| x y \right|}
{\left| y \right| \sqrt{\left| 1 - \left| x \right|^2 \right|}},
\\
r_{\beta \beta} &=&
\frac{\left| y^\ast V_{e1} + x^\ast V_{e2} + x^\ast y^\ast V_{e3} \right|}
{\left| y \right| \sqrt{\left| 1 - \left| x \right|^2 \right|}},
\\
r_{\nu_e} &=&
\frac{\left| y^\ast V_{e1} \right| + \left| x^\ast V_{e2} \right|
+ \left| x^\ast y^\ast V_{e3} \right|}
{\left| y \right| \sqrt{\left| 1 - \left| x \right|^2 \right|}}.
\ea
\es

Numerically,
I have found that the conditions~(\ref{o5}) fit experiment
in the same two types of cases as the conditions~(\ref{o4}).
Thus,
for an inverted neutrino mass spectrum and $\alpha = e$,
one has either $\cos{\delta} \gtrsim 0.4$ for $\beta = \mu$
or $\cos{\delta} \lesssim -0.4$ for $\beta = \tau$.
In both cases
\be
\begin{array}{rcccl}
2.058 &<& r_\mathrm{cosmological} &<& 2.087, \\
0.35 &<& r_{\beta \beta} &<& 0.54, \\
0.977 &<& r_{\nu_e} &<& 0.990.
\end{array}
\ee
On the other hand,
for $\beta = e$ there is another set of cases,
with $\left| \cos{\delta} \right| \lesssim 0.15$,
$r_\mathrm{cosmological} \gtrsim 2$,
and $r_{\beta \beta} \approx r_{\nu_e} \gtrsim 0.5$.
These cases may have $\theta_{23}$ either in the first octant---for a
normal neutrino mass spectrum when $\alpha = \tau$ and for an
inverted neutrino mass spectrum when $\alpha = \mu$---or
in the second octant---interchanging $\alpha = \mu$ with $\alpha = \tau$.
The remarkable feature of this second set of cases
is that the neutrino masses become almost degenerate
when $\theta_{23}$ approaches $\pi/4$.\footnote{In ref.~\cite{Grimus:2006wy}
a model was discussed in which the neutrino masses approach degeneracy
when $s_{12}^2 \to 1/3$,
with $s_{12}^2 < 1/3$ for a normal neutrino mass spectrum
and $s_{12}^2 > 1/3$ for an inverted spectrum.
The present cases display analogous features,
with $s_{12}$ replaced by $s_{23}$ and $1/3$ replaced by $1/2$
just as in ref.~\cite{Grimus:2011sf}.}

\section{Conclusions}

In this paper I have exhaustively classified all the predictive cases where
a type-I see-saw mechanism based on three right-handed neutrinos
has a diagonal charged-lepton mass matrix $M_\ell$
and both the neutrino Dirac mass matrix $M_D$
and the right-handed-neutrino Majorana mass matrix $M_R$ have texture zeros.
Most of the cases with predictive power
had already been studied in the literature,
but I have discovered a few new ones.
The new cases predict either $\left| \cos{\delta} \right| \gtrsim 0.5$
or $\left| \cos{\delta} \right| \lesssim 0.1$;
some of the latter cases feature quasi-degenerate neutrinos
when $\theta_{23}$ is very close to $\pi/4$.

\vspace*{6mm}

\noindent I thank Walter Grimus for reading the manuscript and commenting on it.
This work was supported through the projects PEst-OE-FIS-UI0777-2013,
PTDC/FIS-NUC/0548-2012,
and CERN-FP-123580-2011
of \textit{Funda\c c\~ao para a Ci\^encia e a Tecnologia};
those projects are partially funded through POCTI (FEDER),
COMPETE,
QREN,
and the European Union.

\end{document}